\documentclass[aps,prb,twocolumn,superscriptaddress,showpacs]{revtex4}
\usepackage{graphicx}

\begin{document}
\title{Electron energy spectrum and the Berry phase in graphite
bilayer}

\author{G.~P.~Mikitik}
\affiliation{B.~Verkin Institute for Low Temperature Physics \&
  Engineering, Ukrainian Academy of Sciences,
   Kharkov 61103, Ukraine}

\author{Yu.~V.~Sharlai}
\affiliation{B.~Verkin Institute for Low Temperature Physics \&
  Engineering, Ukrainian Academy of Sciences,
   Kharkov 61103, Ukraine}

\begin{abstract}
We emphasize that there exist four Dirac-type points in the
electron-energy spectrum of a graphite bilayer near the point K of
its Brillouin zone. One of the Dirac points is at the point K, and
three Dirac points lie nearby. Each of these three points
generates the Berry phase $\pi$, while the Dirac point at K gives
the phase $-\pi$. It is these four points that determine the Berry
phase in the bilayer. If an electron orbit surrounds all these
points, the Berry phase is equal to $2\pi$.
\end{abstract}

\date{\today}


\maketitle

In the recent Letter \cite{MF} McCann and Fal'ko considered the
electron-energy spectrum of a graphite bilayer in the vicinity of
the point K of its Brillouin zone and stated that in this bilayer
the low-energy electronic excitations correspond to chiral
quasiparticles with a parabolic dispersion exhibiting Berry phase
$2\pi$. This value of the Berry phase explains the unconventional
Hall effect observed in the bilayer \cite{N1}. Here we refine some
details of the spectrum and show that in reality this Berry phase
in the bilayer is generated by four Dirac points of its spectrum.

 \begin{figure}  
\includegraphics[scale=1.0]{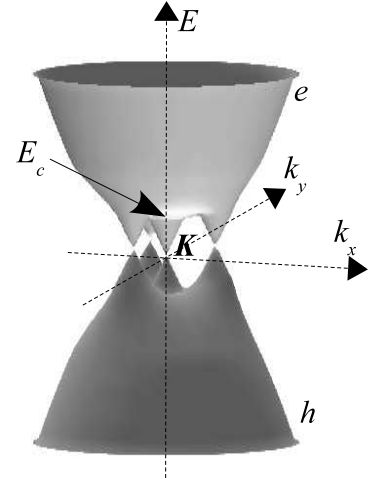}
\caption{\label{fig1} Shown is the dispersion law $E(k_x,k_y)$ of
the electrons (e) and holes (h) in the graphite bilayer near the
point K. The spectrum contains four Dirac points. At the energy
$E_c\sim 2$ meV all the electron cones merge. Note that without
neglecting the parameter $\gamma_4\approx 0.044$ eV, the spectrum
$E(k_x,k_y)$ is slightly asymmetric relative to the plane $E=0$:
The Dirac points in the central and side cones are at energies
$E=0$ and $E=2\gamma_4\gamma_1\gamma_3^2 /\gamma_0^3\approx
0.15E_c$, respectively. Thus, in contrast with graphene, the
graphite bilayer is a semimetal, and changing the Fermi level in
it, one cannot obtain the concentration of the charge carriers
less than $\sim 10^8$ cm$^{-2}$. Scales of the plot are distorted
for clarity.
} \end{figure}   

First of all we point out that the electron spectrum of the
bilayer \cite{MF} can be easily obtained from the well-known
Slonzewski-Weiss-McClure spectrum \cite{SWM} of bulk graphite if
one puts $\cos\xi =0.5$ and $\gamma_2=\gamma_5=0$ (the parameters
$\gamma_2$, $\gamma_5$ describe the interaction of the atoms in
the next-nearest-neighbor layers of graphite that are absent in
the bilayer; $\xi$ is the dimensionless wave vector perpendicular
to the graphite layers). The Slonzewski-Weiss-McClure model
\cite{SWM} describes the wave-vector dependence of four electron
energy bands of graphite $E({\bf k})$ in the vicinity of the
vertical edge HKH of its Brillouin zone. These bands can be found
from the forth-order secular equation:
\begin{equation}\label{1}
\rm{ det} \left | \hat{H} - E \right | =0,
\end{equation}
where the Hamiltonian matrix $\hat{H}$ has the form
\begin{equation}\label{2}
\hat{H}= \left (
    \begin{array}{cccc}
    E_1 & 0 & H_{13} & H_{13}^{*} \\
    0 & E_2 & H_{23} & -H_{23}^{*} \\
    H_{13}^{*} & H_{23}^{*} & E_3 & H_{33} \\
    H_{13} & -H_{23} & H_{33}^{*} & E_3
    \end{array}
\right ).
\end{equation}
Here the following notations have been used:
\begin{eqnarray}
E_1=\Delta+\gamma_1\Gamma+\frac 12 \gamma_5\Gamma^2, \nonumber \\
E_2=\Delta-\gamma_1\Gamma+\frac 12 \gamma_5\Gamma^2, \nonumber \\
E_3=\frac 12 \gamma_2 \Gamma^2, \label{3} \\
 H_{13}=\frac 1{\sqrt{2}}  (
   -\gamma_0+\gamma_4\Gamma
 ) {\rm e}^{i\alpha}\zeta, \nonumber \\
H_{23}=\frac 1{\sqrt{2}} \left (
    \gamma_0+\gamma_4\Gamma
\right ) {\rm e}^{i\alpha} \zeta, \nonumber \\
H_{33}=\gamma_3\Gamma{\rm e}^{i\alpha}\zeta, \nonumber
\end{eqnarray}
where $\alpha$  is the angle between the direction of the vector
${\bf k}$ and the $\Gamma K$ direction in the Brillouin zone;
$\Gamma=2\cos \xi$; $\xi$ and $\zeta$ are dimensionless wave
vectors in the direction of the $z$-axis (i.e. HKH axis) and in
the basal plane, respectively: $\xi=(\pi/ 2)( k_z /|KH|)$,
$\zeta=( 2\pi /\sqrt{3})(k_\bot/|\Gamma K|)$;
$k_\bot=\sqrt{k_x^2+k_y^2}$; ${\bf k}$ is measured from the point
K. The parameter $\gamma_0$ which describes the interaction
between neighbor atoms in a graphite layer is sufficiently large
as compared to the other parameters $\gamma_i$, $\Delta$ which
describe relatively weak interactions between atoms in different
graphite layers. \cite{c1} As it was mentioned above, the spectrum
of the bilayer is obtained if one puts $\Gamma =1$ and
$\gamma_2=\gamma_5=0$. Note that in this way one can allow for the
small parameter $\gamma_4$ that was neglected in
Ref.~\onlinecite{MF}.

In the interval $E_c \ll |E| \ll \gamma_1$ equations
(\ref{1})-(\ref{3}) lead to the approximate formula for the two
low-energy bands of electrons (e) and holes (h) in the bilayer
 \begin{equation}\label{4}
 E^{e,h}(k_x,k_y) \approx \pm{\gamma_0^2 \over \gamma_1}
 \zeta^2,
 \end{equation}
that exhibits a quadratic dependence on $k_{\bot}$ discussed in
Ref.~\onlinecite{MF}. Here $E_c= (\gamma_1/4)
(\gamma_3/\gamma_0)^2\approx 2$ meV, and the signs plus and minus
correspond to the electrons and holes, respectively.

At energies $|E|<E_c$ the role of the so-called trigonal warping
\cite{c} increases, and this warping breaks the line
$E=\,$constant in the $k_x$-$k_y$ plane into one central and three
side pockets, \cite{MF} see Fig.~2 in Ref.~\onlinecite{MF}. We
emphasize here that each of these pockets contains a point at
which the electron and the hole bands contact, and near all these
points the spectrum is {\it linear} in ${\bf k}$, Fig.~1, see also
Refs.~\onlinecite{KA,PP,MGV}. Thus, similarly to bulk graphite
\cite{G} and in contrast with graphene, in the graphite bilayer
near the point K there are {\it four} points of the Dirac type.
The central contact point coincides with K, while the three side
contact points are at a distance of $(\sqrt 3/
2\pi)(\gamma_3\gamma_1/ \gamma_0^2) |\Gamma K|\approx 0.005|\Gamma
K|$ from K where $|\Gamma K|$ is the distance between the point K
and the center $\Gamma$ of the Brillouin zone.

Before analyzing the Berry phase in the bilayer, we point out a
general property of this phase for closed semiclassical electron
orbits in crystals with inversion symmetry in the magnetic field
$H$. The Berry phase $\Phi_B$ for such an orbit $\Gamma$ lying in
the Brillouin zone and belonging to a band $0$ is given by
\cite{Zak,MS}
\begin{equation}\label{5}
\Phi_B= \oint_{\Gamma} {\bf \Omega} d {\bf k},
\end{equation}
where the direction of the integration is determined by the vector
$[{\bf H}\times \partial E^0/\partial {\bf k}]$,  ${\bf \Omega}$
is the intraband matrix element of the periodic (in ${\bf k}$)
part of the coordinate operator in the crystal momentum
representation, \cite{Bl}
\begin{equation}\label{5a}
{{\bf \Omega}({\bf k})=\imath \int d{\bf r} u_{{\bf
k}0}^{\ast}({\bf r}){\bf {\nabla}_k } u_{{\bf k}0}}({\bf r}),
\end{equation}
and ${u_{{\bf k}0}({\bf r})}$ is the periodic factor in the
electron Bloch wave function of the band $0$, ${{\psi}_{{\bf
k}0}({\bf r})=\exp (\imath {\bf k r}) u_{{\bf k}0}({\bf r})}$.
This Berry phase manifests itself in the Onzager-Lifshitz-Kosevich
quantization condition \cite{Ons,LK} for energy levels
$\varepsilon$ of an electron in the magnetic field,
 \begin{equation}\label{6}
 S(\varepsilon ,k_H)=\frac{2\pi e H}{\hbar c}(n+\gamma ),
 \end{equation}
where  ${S}$ is the cross-sectional area of the closed orbit in
the ${\bf k}$ space; $k_H$ is the component of ${\bf k}$ along the
magnetic field ${\bf H}$; ${n}$ is a large integer (${n>0}$);
${e}$ is the absolute value of the electron charge, and the
constant $\gamma$ is given by the formula:\cite{MS}
 \begin{equation}\label{7}
 \gamma = \frac{1}{2}-\frac{\Phi_B}{2\pi}\ .
 \end{equation}
If one reverses the direction of the magnetic field, the direction
of the integration in Eq.~(\ref{5}) is also reversed, and the
Berry phase changes its sign, while $\gamma$ changes by
$\Phi_B/\pi$. But the electron spectrum in the magnetic field has
to be invariant under this transformation. Since in the
semiclassical approximation the quantity $\gamma$ is defined up to
an integer, we conclude that the Berry phases for such the orbits
are always multiple of $\pi$. \cite{Zh} This property of the Berry
phase agrees with the results of Ref.~\onlinecite{MS}. It was
shown in that paper that if the electron orbit $\Gamma$ surrounds
band-contact lines (Dirac points in the two dimensional case),
each of the lines (the Dirac points) contributes $\pm \pi$ to
$\Phi_B$. If the orbit does not surround the band-contact line
(the Dirac points), the Berry phase is equal to zero. This
property of the Berry phase also means that $\Phi_B$ can change
only abruptly when the crystal potential is perturbed, and a small
variation of this potential can, in principle, lead to an
essential change of the Berry phase.

McCann and Fal'ko derived the Berry phase for the electron orbits
in the bilayer from an effective Hamiltonian that leads to the
parabolic spectrum (\ref{4}). However, it is clear that parabolic
spectra are idealization, and they never occur in crystals, and at
least a small warping of these spectra always exists. As it was
mentioned above, this small warping may essentially change the
Berry phase, and in principle, different symmetries of the small
warping may lead to different values of the Berry phase since the
number of the Dirac points depends on symmetry of a crystal. In
the case of the bilayer the trigonal warping generates the three
additional side Dirac points and changes the type of the central
band-contact point. Thus, for the derivation of the Berry-phase
value to be justified, it is necessary to consider the real
symmetry of the spectrum in the bilayer and to take into account
all the band-contact points of this spectrum.

Although each of the four Dirac point in the bilayer generates the
Berry phase $\pm \pi$, it is necessary to find the signs of these
phases. This can be done using the approach of
Ref.~\onlinecite{Zh} in which the effect of a small spin-orbit
interaction on the Berry phase was investigated. It is important
that if this interaction is weak, it does not change the Berry
phase. But the interaction enables one to fix the sign of the
phase. \cite{c3} This is clear from the following considerations:
Without the spin-orbit interaction the quantity ${ [ {\bf
{\nabla}_k} \times {\bf \Omega} ] }$ is singular at the Dirac
points. The interaction lifts the band degeneracy, and $[ {\bf
{\nabla}_k} \times {\bf \Omega} ]$ becomes a smooth function which
can be calculated using the formula \cite{Bl} [see also
expressions (9), (10) in Ref.~\onlinecite{Ber} and formula (6.6)
in Ref.~\onlinecite{Niu}],
 \begin{equation} \label{8}
[ {\bf {\nabla}_k} \times {\bf \Omega} ]_z=i \hbar^2 \sum_{l \neq
0} \frac {v^{x}_{0l}v^{y}_{l0}- v^{y}_{0l}v^{x}_{l0}} {\left (
E_l({\bf k}) - E_0({\bf k}) \right )^2 },
 \end{equation}
that is completely independent of a gauge of the electron Bloch
wave functions. Here $v^{i}_{0l}$ are iterband matrix elements of
the velocity operator for the bands $0$ and $l$ at a point ${\bf
k}$. Taking into account that Eq.~(\ref{5}) may be rewritten in
the form
 \begin{equation}\label{9}
\Phi_B=\int_S dk_x dk_y[{\bf {\nabla}_k} \times {\bf \Omega}]_z ,
 \end{equation}
where the integration is over the surface $S$ spanning the orbit
$\Gamma$, one can unambiguously calculate the Berry phase for each
of the Dirac points in the bilayer.

 \begin{figure}  
\includegraphics[scale=1.0]{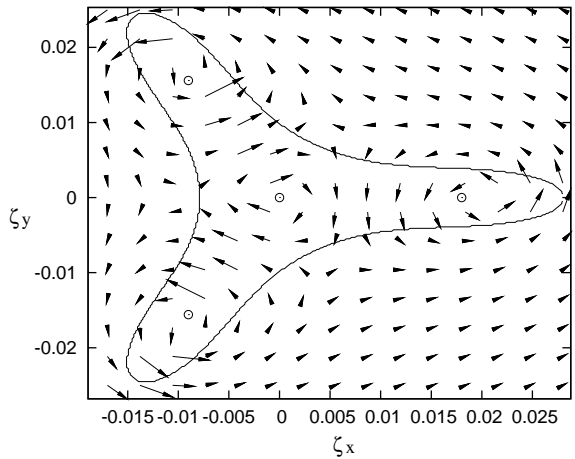}
\caption{\label{fig2} The field ${\bf \Omega}(k_x,k_y)$ (arrows)
for the electron band e in the bilayer. The length of the arrows
is proportional to $|{\bf \Omega}|$. Small circles mark the
positions of the Dirac points in the $k_x$-$k_y$ plane, and the
solid line shows an electron orbit with $E>E_c$. Note that near
the central and the side Dirac points the vector ${\bf \Omega}$
``circulates'' in opposite directions. The components $k_i$ are
given in the dimensionless units $\zeta_i=( 2\pi
/\sqrt{3})(k_i/|\Gamma K|)$ where $|\Gamma K|$ is the distance
between the point K and the center of the Brillouin zone,
$\Gamma$.
} \end{figure}   

The effect of the spin-orbit interaction on the spectrum of
graphene was considered in Refs.~\onlinecite{KM,Sin}, and the
following Hamiltonian for the electron and hole bands in the
vicinity of the point K was obtained
 \begin{equation}\label{10}
\hat H=v(k_x\sigma_x+k_y\sigma_y) + \Delta_{so}\sigma_z s_z ,
 \end{equation}
where $\sigma_i$ are the Pauli matrices describing the two bands,
$v$ is a matrix element proportional to $\gamma_0$,
$2\Delta_{so}\sim 0.2$ meV is the gap due to the spin-orbit
interaction, and $s_z$ is the Pauli matrix representing the
electron's spin. Since the interaction between carbon atoms in a
graphite layer is larger than the interaction between the atoms in
different graphite layers (i.e. $\gamma_0\gg \gamma_i$), we use
Hamiltonian (\ref{10}) of a single layer to include the spin-orbit
interaction in Hamiltonian (\ref{2}), (\ref{3}) of the bilayer.
Then, in the leading order in the small parameter
$\Delta_{so}/\gamma_1$ the matrix elements $E_3$ in the third and
forth lines of formula (\ref{2}) should be replaced by $E_3-
\Delta_{so}$ and $E_3 + \Delta_{so}$, respectively.

Although formulas (\ref{8}), (\ref{9}) clearly demonstrate the
invariance of the Berry phase relative to the gauge
transformations of the Bloch wave functions, it is more convenient
to find the field ${\bf \Omega}(k_x,k_y)$ in the bilayer, and to
calculate the Berry phase directly from formula (\ref{5}). This
field for a band $0$ can be obtained with the formula,
\cite{Zh,c2}
 \begin{equation}\label{11}
 {\bf \Omega} = i\left ( S^+{\partial S\over \partial {\bf k}}
 \right )_{\!\!00},
 \end{equation}
where $S({\bf k})$ is the matrix reducing Hamiltonian (\ref{2}),
(\ref{3}) to the diagonal form, $S^+$ is the Hermitian conjugate
matrix, and the subscript means that one has to consider the
diagonal matrix element corresponding to the band $0$. Assuming
the spin-orbit interaction is infinitesimal, a direct calculation
of $S$ leads to the field ${\bf \Omega}(k_x,k_y)$ shown in Fig.~2.
Note that near the central and the side Dirac points the vector
${\bf \Omega}$ ``circulates'' in opposite directions, which means
the opposite signs of the Berry phases generated by these points.
The calculation of the integral (\ref{5}) gives the Berry phase
$\pi$ for electron orbits surrounding each of the three side Dirac
points and $-\pi$ for an orbit around the central point. Since the
Berry phase of an electron does not depend on a size or a shape of
its orbit in the Brillouin zone but is determined only by the
Dirac points enclosed by the orbit, \cite{MS} one finds the Berry
phase $3\cdot\pi -\pi= 2\pi$ for electrons with energies $E>E_c$
(i.e., at the electron concentration $N>N_c\approx 5\cdot 10^{10}$
cm$^{-2}$) when their orbits surround all these four points. Thus,
at $E>E_c$ we arrive at the same Berry phase $2\pi$ as in the case
of the parabolic spectrum (\ref{4}) in spite of the change in the
energy-band degeneracy. The difference in the Berry phases for the
parabolic and real spectra will manifest itself only in the
interval $-E_c<E<E_c$ at low magnetic fields when many Landau
levels lie in this interval, and hence the trigonal warping is not
a small perturbation as compared to the magnetic energy of the
electron.

The above considerations also clarify existence of the resistivity
maximum discovered in the bilayer. \cite{N1} In graphene the
universal resistivity maximum $h/4e^2$ was observed at zero
magnetic field and low charge-carrier concentration $N$, and this
maximum was explained by absence of localization for electrons
with the Dirac-type spectrum. \cite{N2} A similar resistivity
maximum was also observed in the bilayer, \cite{N1} and Novoselov
{\it et al}. \cite{N1} emphasized that this observation is
unexpected due to the parabolic spectrum in the bilayer. Although
the parabolic spectrum of Ref.~\onlinecite{MF} and the spectrum
with the four Dirac points lead to the same Berry phase at
$N>N_c$, existence of the Dirac points seems to shed a light on
appearance of the resistivity maximum in the bilayer. The effect
of the trigonal warping on the resistivity maximum in the bilayer
was quantitatively analyzed in Refs.~\onlinecite{KA,CCD}.

Finally, let us point out that our result for the Berry phase is
reminiscent of the result of Ref.~\onlinecite{MGV}. Authors of
that paper studied stability of Fermi points in multilayer
graphene relative to perturbations of the crystal potential, and
they found that in the bilayer some topological charge Q is equal
to 2 for the case of the parabolic spectrum (\ref{4}), while if
one takes into account the trigonal warping, each of the three
side Dirac points located near the point K has the charge 1 and
the charge of the central Dirac point at K is $-1$. Nevertheless,
it is necessary to emphasize that in general case formula (11)
defining Q in Ref.~\onlinecite{MGV} does not coincide with
$\Phi_B/\pi$ [compare this formula with our Eqs.~(\ref{5}) and
(\ref{11})].


\begin{thebibliography}{}
\bibitem{c1} Below we use the same values of the parameters as in
Ref.~\onlinecite{MF}: $\gamma_0=2.47$ eV, $\gamma_1=0.39$ eV,
$\gamma_3=0.315$ eV, $\gamma_2=\gamma_5=\gamma_4=\Delta=0$.
\bibitem{c} The trigonal warping is due to the term proportional
to $\gamma_3$ in Hamiltonian (\ref{2}), (\ref{3}).
\end{thebibliography}
\end{document}